\documentclass[8.5pt,twoside,twocolumn]{article}
\usepackage[super,sort&compress,comma]{natbib} 
\usepackage{mhchem}
\usepackage{times,mathptmx}
\usepackage{sectsty}
\usepackage{balance} 

\usepackage{graphicx} 
\usepackage{lastpage}
\usepackage[format=plain,justification=raggedright,singlelinecheck=false,font=small,labelfont=bf,labelsep=space]{caption} 
\usepackage{fancyhdr}
\begin{document}

\thispagestyle{plain}
\fancypagestyle{plain}{
\renewcommand{\headrulewidth}{1pt}}
\renewcommand{\thefootnote}{\fnsymbol{footnote}}
\renewcommand\footnoterule{\vspace*{1pt}%
\hrule width 3.4in height 0.4pt \vspace*{5pt}} 
\setcounter{secnumdepth}{5}

\makeatletter 
\def\subsubsection{\@startsection{subsubsection}{3}{10pt}{-1.25ex plus -1ex minus -.1ex}{0ex plus 0ex}{\normalsize\bf}} 
\def\paragraph{\@startsection{paragraph}{4}{10pt}{-1.25ex plus -1ex minus -.1ex}{0ex plus 0ex}{\normalsize\textit}} 
\renewcommand\@biblabel[1]{#1}            
\renewcommand\@makefntext[1]%
{\noindent\makebox[0pt][r]{\@thefnmark\,}#1}
\makeatother 
\renewcommand{\figurename}{\small{Fig.}~}
\sectionfont{\large}
\subsectionfont{\normalsize} 

\fancyfoot{}
\fancyhead{}
\renewcommand{\headrulewidth}{1pt} 
\renewcommand{\footrulewidth}{1pt}
\setlength{\arrayrulewidth}{1pt}
\setlength{\columnsep}{6.5mm}
\setlength\bibsep{1pt}

\twocolumn[
  \begin{@twocolumnfalse}
\noindent\LARGE{\textbf{Nature of proton transport in a water-filled carbon nanotube and in liquid water$^\dag$}}
\vspace{0.6cm}

\noindent\large{\textbf{Ji Chen,\textit{$^{a}$} Xin-Zheng Li,$^{\ast}$\textit{$^{b}$}\textit{$^{c}$} Qianfan Zhang,\textit{$^{b}$} Angelos Michaelides,\textit{$^{c}$} and Enge Wang,$^{\ast}$\textit{$^{a}$}}}\vspace{0.5cm}



\noindent \normalsize{Proton transport (PT) in bulk liquid water and within a thin water-filled carbon nanotube
has been examined with \textit{ab initio} path-integral molecular dynamics~(PIMD).
Barrierless proton transfer is observed in each case 
when quantum nuclear effects~(QNEs) are accounted for. 
The key difference between the two systems is that in the nanotube facile PT is facilitated by
a favorable pre-alignment of water molecules,
whereas in bulk liquid water solvent reorganization
is required prior to PT. Configurations where the quantum excess proton is delocalized
over several adjacent water molecules along with continuous interconversion between
different hydration states reveals that, as in liquid water, the hydrated proton under
confinement is best described as a fluxional defect, rather than any individual 
idealized hydration
state such as Zundel, Eigen, or the so-called linear H$_7$O$_3^+$ complex along the water chain. 
These findings highlight the importance of QNEs in intermediate
strength hydrogen bonds~(HBs) and explain why H$^+$ diffusion through nanochannels is impeded 
much less than other cations.
}
\vspace{0.5cm}
 \end{@twocolumnfalse}

]

\section{Introduction}


\footnotetext{\textit{$^{a}$~ICQM and School of Physics, Peking University, Beijing 100871, P. R. China}}
\footnotetext{\textit{$^{b}$~School of Physics, Peking University, Beijing 100871, P. R. China}}
\footnotetext{\textit{$^{b}$~Thomas Young Centre, London Centre for Nanotechnology and Department of Chemistry, University College London, London WC1E 6BT, U.K.}}



Proton transport (PT) under confinement
is an elementary chemical process of central importance to biology 
through e.g. enzyme catalysis and membrane transport.\cite{marxnat, voth, dellago}
It is also central to technological processes such as
electrical
power generation in hydrogen fuel cells\cite{umeyama} and emerging applications 
in nanotechnology involving the controlled flow of protons within carbon
nanotubes~(CNT).\cite{snow, baughman2002}
Motivated by this broad fundamental and technological importance, a large
number of experimental and theoretical studies have been performed in order to understand
PT within hydrophobic channels such as CNTs, aquaporin~(AQP), and gramicidin 
A~(GA).\cite{dellago, brewer, klein, cukierman, LiAQP, jensen, hoarfrost, bureekaew}
The most commonly used experimental approach for exploring PT under confinement
involves electrophysiology 
techniques, where
current-voltage relations of channels are measured for different concentrations of protons
and other ions.\cite{cukierman}
Recently, 
nuclear magnetic resonance 
and quasi-elastic neutron scattering (QENS) have also
been employed to study the local dynamic change of protons under 
confinement.\cite{hoarfrost, bureekaew}
Although these methods are powerful for understanding the behavior of protons and 
ionic conductivity in various environments, interpretation of the data beyond a
qualitative level is difficult.

Molecular dynamics (MD) simulations, on the other hand, are a powerful tool for explaining
details of processes such as PT at the atomic scale.
Various earlier studies have demonstrated that in aqueous systems protons hop from one water 
molecule to the next via a Grotthuss mechanism, resulting in transport
of charge defects rather than individual atoms.\cite{voth, voth2006, marxnat}
Using the multistate empirical
valence bond~(MS-EVB) method along with a classical description of the nuclei, Brewer~\textit{et~al.}\cite{brewer} and
Dellago~\textit{et~al.}\cite{dellago} 
have carried out pioneering studies of PT in a CNT. 
Exceedingly fast PT was reported, $\sim$40 times its value in bulk water.\cite{dellago}
This prediction is consistent with the experimental observation that the ratio of
H$^+$ to other cation's (e.g. Na$^+$, K$^+$) diffusion coefficients increases 
upon nanoconfinement.\cite{pomes,roux}
This rapid diffusion was attributed
to a decrease of the classical proton transfer free-energy barrier 
compared with that in bulk water.\cite{brewer}
However, in an alternative explanation it has been suggested that it is the local conformation
of HBs between the water molecules 
within the CNT 
that facilitates rapid PT under confinement.\cite{Koefinger}
Both explanations are compelling and so the fundamental mechanism of PT under the
seemingly simple conditions of confinement within CNTs remains
in doubt. 

From a theoretical perspective, the most effective way of distinguishing between the two models is to perform
simulations that directly compare the mechanism of PT in bulk water and in a CNT.
Ideally this would also involve an approach that accounts for the quantum nature of the nuclear
degrees of freedom in addition to the electronic degrees of freedom, since 
it is well known that QNEs~(quantum delocalization, 
tunnelling, and zero point energy)
are critical to PT. ~\cite{marxnat,tuckermannat, schmitt1999,xzli,xzl2}
%
To this end we present in the current study an extensive series 
of \textit{ab initio} MD
and \textit{ab initio} PIMD simulations
for PT in bulk liquid water and within a thin water-filled CNT. 
With \textit{ab initio} MD, bond making and bond breaking events
as well as thermal (finite temperature) effects can be accounted for in a seamless
manner based on the forces computed ``on the fly'' as the dynamics of the system
evolves. 
By going beyond this with \textit{ab initio} PIMD quantum effects of the nuclei
are also accounted for and by 
comparing the results
obtained from MD and PIMD the role of QNEs can be examined in a very clean 
manner.\cite{berne1993,parrinello1996,marxnat,xzli,xzl2,zhang}

Despite several excellent previous studies, this is the first \textit{ab initio}
PIMD report of PT in a water-filled CNT and consequently the first direct
comparison with PT in bulk water. 
The key findings of this study are: i) barrierless proton transfer
is observed in bulk water and within the
nanotube when QNEs are taken in to account;
ii) the biggest difference between the two systems is that a favorable pre-alignment 
of water molecules
within the tube facilitates the facile PT, whereas in the bulk liquid solvent reorganization
is required prior to PT; and iii) there is a continuous interconversion between different hydration
states, namely Zundel, Eigen (or the so-called linear H$_7$O$_3^+$ complex along the 1D water chain
\cite{voth}), and configurations 
when the quantum excess proton is delocalized over several adjacent water molecules, which leaves
the excess proton best described
as a fluxional defect\cite{marxnat} 
in both systems.
These findings emphasize the importance of an accurate account of both electronic and nuclear degrees of
freedom when describing PT and explain
in a simple manner why protons move much less sluggishly through nanochannels than other cations do.

\section{Results}

\begin{figure}[h]
\centering
  \includegraphics[width=0.45\textwidth]{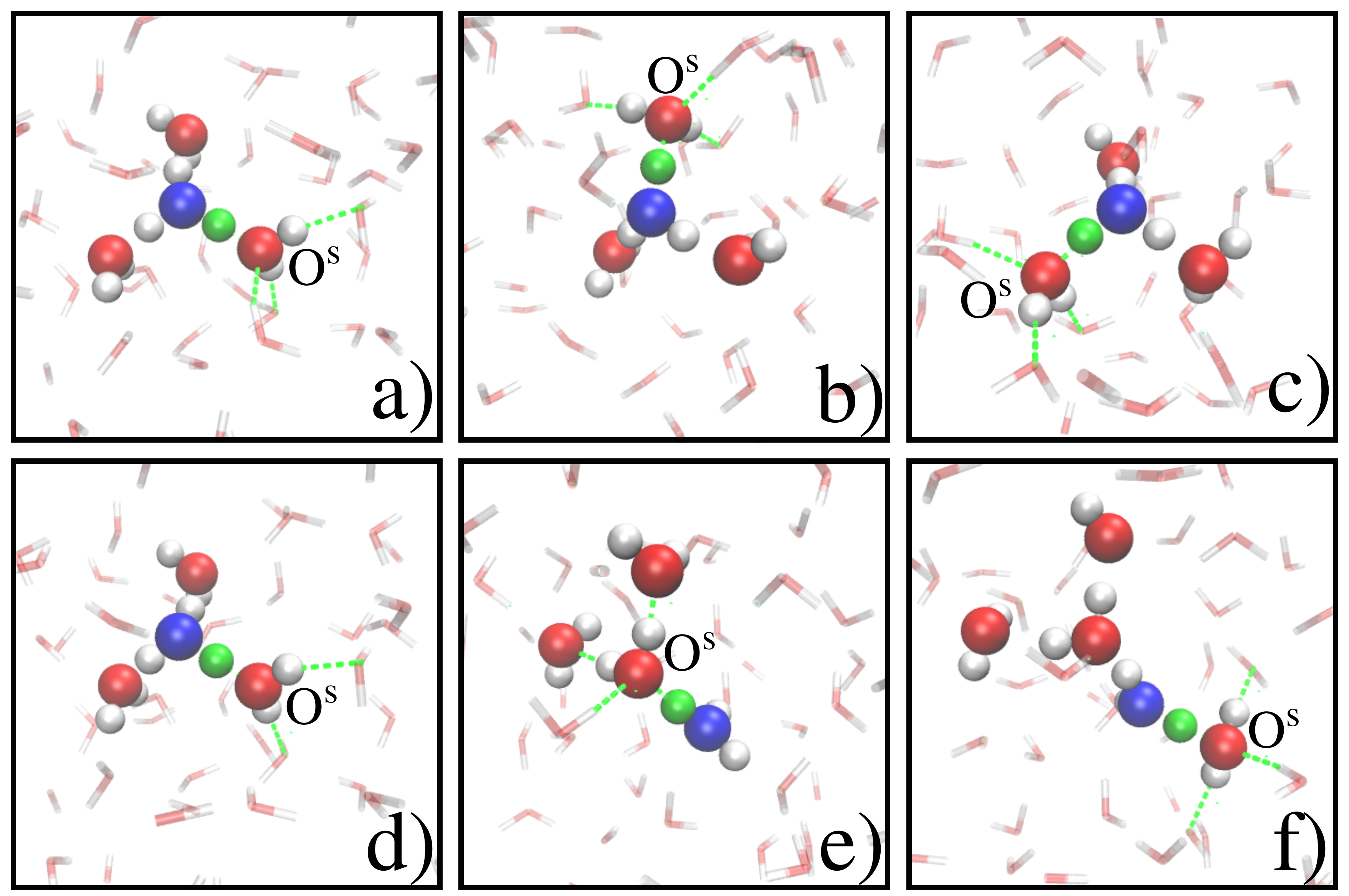}
  \caption{Snapshots of PT, focusing on the active region, in an \textit{ab initio} MD simulation
of bulk liquid water with classical nuclei. 
Red spheres are oxygen atoms and white spheres are hydrogen atoms.
The excess proton is in green and the pivot oxygen (O$^{\text{p}}$) 
is in blue. O$^{\text{s}}$ is the oxygen which shares the most active proton with the pivot oxygen.
To illustrate the effect of solvent reorganization, besides the most active 
hydrogen bond~(HB), other
HBs associated with O$^{\text{s}}$ are indicated by green dashed lines.
Panels (a), (b), and (c)
show that the most active proton switches identity between the
three HBs related to O$^{\text{p}}$, where the coordinate number of O$^{\text{s}}$ is 4.
Going from panels (c) to (d), the coordination number of O$^{\text{s}}$ reduces
to 3 and PT occurs.
After PT the most active proton begins to exchange sites about a new pivot oxygen (panels e and f)
and the coordination number of O$^{\text{s}}$ changes back to 4.
}
  \label{figure1}
\end{figure}

We start by discussing the results with classical nuclei in bulk 
liquid water. 
The first quantity introduced is the proton transfer coordinate
$\delta$, which is the difference
in distance between each proton and its two nearest oxygen atoms. 
When $\delta$ is zero for a particular proton that proton is equidistant from
its two nearest oxygens, when the absolute value of $\delta$
is large the proton belongs to one of the two oxygen atoms. 
Further, as in some earlier studies,\cite{marxnat, voth2006}
we define the proton with the smallest $|\delta|$ (otherwise
known as the most active proton) as the excess 
proton (green atom in \ref{figure1}).
Along with the excess proton, we also identify the pivot oxygen~(O$^{\text{p}}$).
This is done by assigning each proton to its closest oxygen and then identifying the
pivot oxygen as the one which at each time step has three protons~(blue atom in \ref{figure1}).
Our simulations of PT in bulk water support the picture 
obtained previously from MD simulations using both MS-EVB~\cite{voth2006,schmitt1999,day2000} and
\textit{ab initio} methods.\cite{tuckerman1995b,marxnat,tuckerman2009}
Essentially it involves the exchange of the most active proton between different
HBs associated with O$^{\text{p}}$ until a suitably under-coordinated water molecule becomes
available to accept the most active proton.
This is shown in \ref{figure1}~(a)-(c) where O$^{\text{s}}$ (the oxygen which 
shares the excess proton with O$^{\text{p}}$)
changes identity until 
a proper under-coordinated most active proton acceptor 
appears (see \ref{figure1}~(d), the coordination number of
O$^{\text{s}}$ has changed from 4 to 3).
After PT from one oxygen to another (\ref{figure1}~(e)), the HB involving the most active proton
begins to switch again (\ref{figure1}~(f)).
The slowest step in this process is the solvent reorganization which is required for a suitably
under-coordinated most active proton acceptor to become 
available (O$^{\text{s}}$ in \ref{figure1}~(d)).

\begin{figure}[h]
\centering
\includegraphics[width=0.45\textwidth]{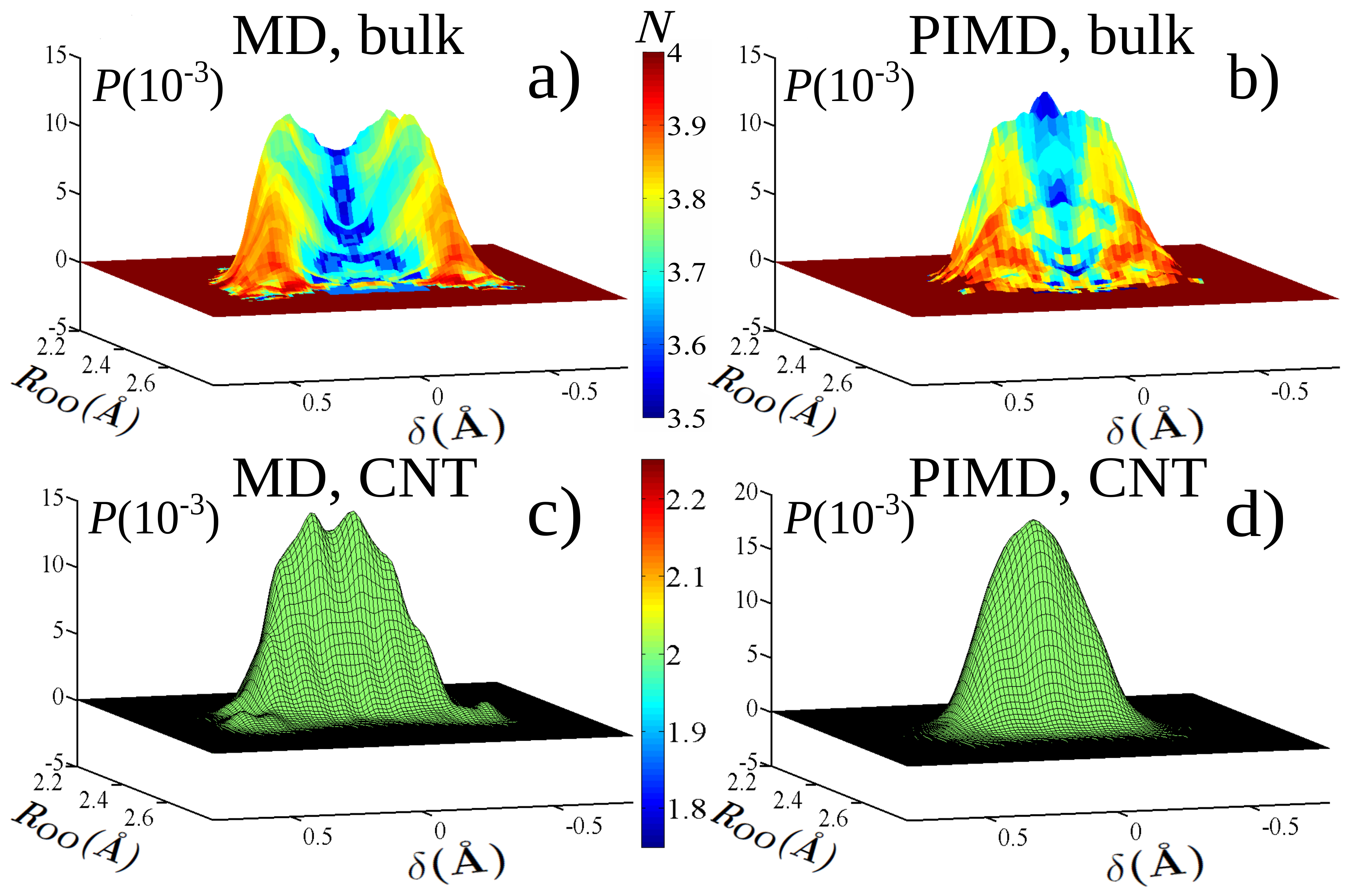}
\caption{\label{figure2}
Probability distributions ($P$) for the hydration state of the most active proton
as a function of $\delta$ and its corresponding $R_{\text{OO}}$. The coordination
number $N$ for the acceptor of the excess proton (O$^\text{s}$), indicated with color,
is superimposed on $P$.
Panel a (b) corresponds to the MD (PIMD) simulation of the excess proton
in bulk liquid water (labeled bulk). Here it can be seen that as $\delta$ 
decreases $N$ also decreases from 4 (red) to 3.5 (blue).
This is consistent with what has been reported by Marx \textit{et al},~\cite{marxnat} 
where same type of figure was also used. 
Panel c~(d) corresponds to the MD (PIMD) simulation of the excess
proton within the water filled carbon nanotube (labeled CNT). In this case 
$N$ remains at about 2 for all values of $\delta$. Therefore, the superimposed
color doesn't change at all.  
}
\end{figure}

To characterize the PT process in more detail, we 
plot the distribution $P$ of the configurations visited during the simulation
as a function of $\delta$ and $R_{\text{OO}}$ (HB length) for the excess proton.
This is represented by the height of the peaks shown in \ref{figure2}~(a).
Along with $P$, we also assess the role of solvent reorganization on O$^{\text{s}}$, 
by looking at its coordination number $N$, \textit{i.e.,} the number of oxygen atoms within
a certain distance (3.2 \AA\text{}) of O$^{\text{s}}$. 
$N$ has been calculated as a function of
$\delta$ and $R_{\text{OO}}$ and is superimposed in color on $P$ in \ref{figure2}~(a).
It can be seen from \ref{figure2}~(a) that at large $|\delta|$ $N$ tends to be around 4 and as
$|\delta|$ reduces so does $N$.
This supports the view that the excess proton is more likely to transfer when
a HB in the second solvation shell of the excess proton is broken,
that is when there is a proper under-coordinated most active proton 
acceptor.
The reduced probability at $\delta$ =  0 indicates that the most 
active proton feels a barrier upon
transferring from one oxygen atom to another. 
As such the corresponding hydration state of the proton
is the Eigen 
complex~(H$_9$O$_4^+$),\cite{day2000,markovitch2008,tuckerman1995b,tuckerman2009,marxnat,zhipan2012}
where the excess proton belongs to O$^{\text{p}}$ instead of being equally shared by
O$^{\text{p}}$ and O$^{\text{s}}$ (Zundel complex, H$_5$O$_2^+$).
 
Let us now look at the impact of QNEs on PT in the bulk. 
This can be seen by comparing 
\ref{figure2}~(a) with the equivalent probability distribution obtained from
PIMD simulations shown in \ref{figure2}~(b). 
The key difference is that the
clear double-peak structure of $P$ in \ref{figure2}~(a) is absent in \ref{figure2}~(b).
This is a consequence of the excess proton's zero-point energy essentially washing
out the classical PT barrier.
The maximum at $\delta=0$ of the peak in \ref{figure2} (b) indicates that Zundel 
complexes (H$_5$O$_2^+$) are enhanced.\cite{marxnat}
However, aside from these differences a key point of similarity is that
the coordination, $N$, decreases as $\delta$ 
approaches zero. 
Since in this case when QNEs are accounted for there is essentially no proton transfer barrier, 
this is a clear indication that HB breaking in the second solvation shell of the
excess proton limits PT.\cite{marxnat}

Moving to PT in the the nanotube, we again plot $P$ and $N$ for the excess proton as 
a function of $\delta$ and 
$R_{\text{OO}}$ (\ref{figure2} (c) and (d)).
In the MD simulation with classical nuclei,
the double-peak structure observed in liquid water is less pronounced. This is consistent with 
the suggestion that
the most active proton feels a smaller classical free energy barrier when constrained 
within the CNT.\cite{brewer}
However, in the PIMD simulation, this free energy barrier 
disappears, revealing that in both the liquid and the nanotube there is no free energy barrier
for proton transfer when QNEs are accounted for. 
Furthermore, if we compare the
upper and lower panels in \ref{figure2}, the most striking difference between the
liquid and the nanotube is
that there is no solvent reorganization for PT within the nanotube.
The coordination number, $N$, is essentially insensitive to $\delta$ and remains at $\sim$2 in both \ref{figure2}~(c) and (d).
Given that in both the bulk and the CNT there is no proton transfer
barrier when QNEs are accounted for, with
solvent reorganization
limiting PT in the bulk, it is clear that the reduced coordination of the molecules in the CNT is 
the key to facile PT in the CNT.
Essentially in the nanotube the water molecules are pre-aligned in a favorable orientation for
proton transfer and solvent reorganization is not required to facilitate this.

\begin{figure}[h]
\centering
\includegraphics[width=0.4\textwidth]{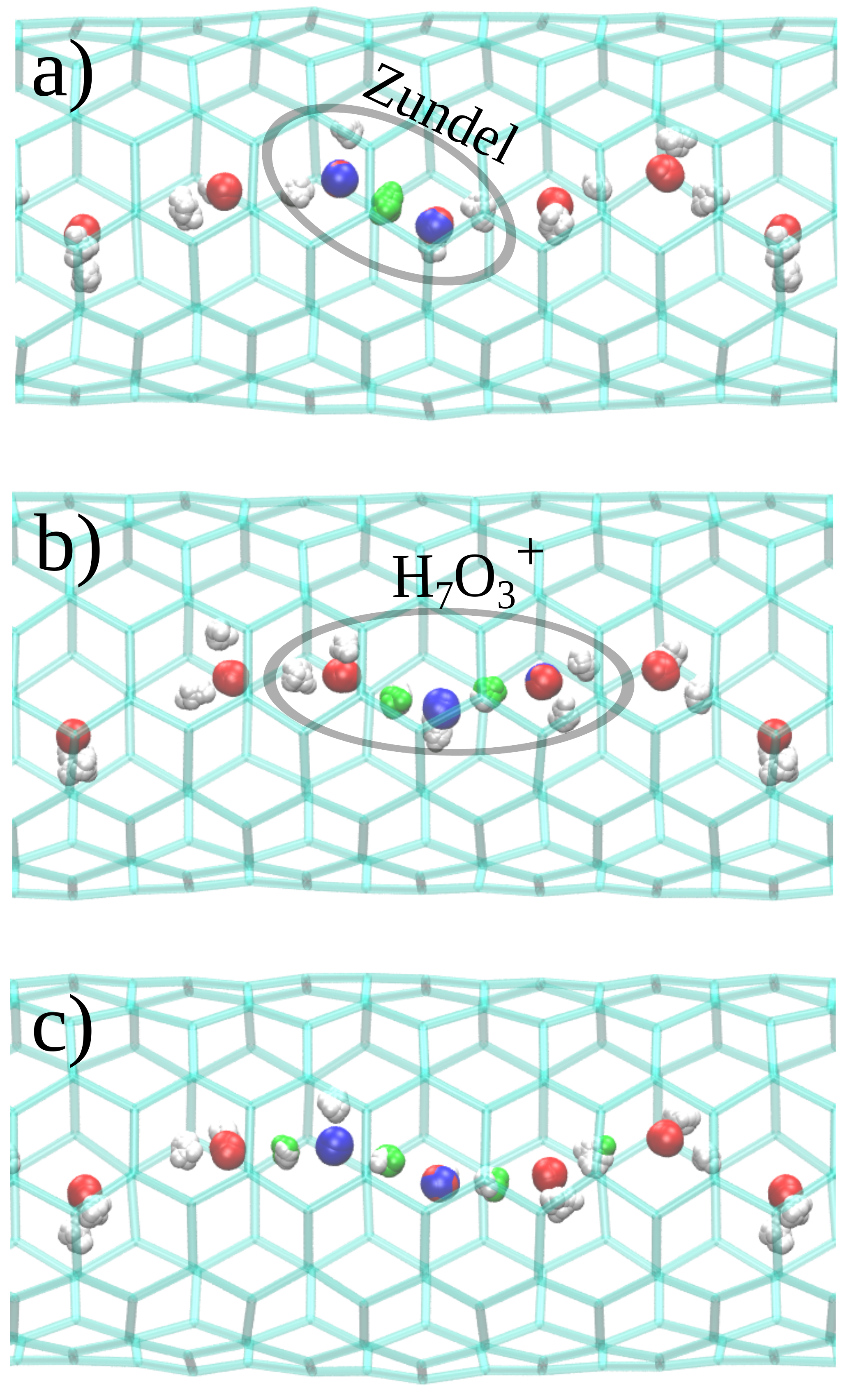}
\caption{\label{figure3}
Three typical snapshots from the PIMD simulation of an excess proton along a water chain
in a CNT. 
In (a) an ideal Zundel~(H$_5$O$_2^+$) is present, in (b) an ideal linear H$_7$O$_3^+$
is present, and in (c) the excess proton is delocalized over several HBs.
As in \ref{figure1} the pivot oxygen is plotted in blue and the most active 
proton is plotted in green.
Other oxygens (protons) are shown in red (white).
In a PIMD simulation, the system is represented by $n$ classical replicas (in this
case $n$ = 32) and as a result the most active proton or pivot oxygen does not 
have to be at the same site at each timestep. This is the case in (a) where the
pivot oxygen is located at two adjacent oxygens, in (b) where the most active proton
is at two sites, and in (c) where the most active proton and pivot oxygen is 
delocalized over several sites. 
}
\end{figure}


%
In a PIMD simulation the system and each atom within it is represented by $n$ classical replicas.
This description of the system in terms of $n$ replicas leads to a
blurring of the atomic positions precisely due to QNEs. 
Further, for each replica there is one most active proton
and one pivot oxygen at each timestep and in each replica
these may not necessarily be the same atom.
Such a scenario can be seen for example in \ref{figure3}~(a) where at the particular 
snapshot shown the pivot oxygen is 
delocalized over two neighboring oxygen sites. 
In this particular state the most
active proton is localized equidistant between the two oxygens and so the state corresponds 
to a Zundel complex H$_5$O$_2^+$. 
\ref{figure3}~(b) shows a snapshot for a different state where the pivot oxygen is mainly localized on one
oxygen atom but the most active proton delocalizes over two neighboring HBs, 
corresponding to a linear H$_7$O$_3^+$ complex observed as the transcient intermediate 
in Ref. \cite{voth}.
Aside from these two conventional states, there are many instances during the PIMD simulation
in which the 
most active proton delocalizes over 3 or 4 HBs along the water chain. 
A snapshot of one such situation is shown \ref{figure3}~(c) where the most active 
proton extends across 4 HBs.
The quantum delocalized state is obviously not found in the MD simulation with classical nuclei and
it reveals the fundamentally different nature of the system when explored with classical and quantum nuclei.

\begin{figure}[h]
\centering
\includegraphics[width=0.45\textwidth]{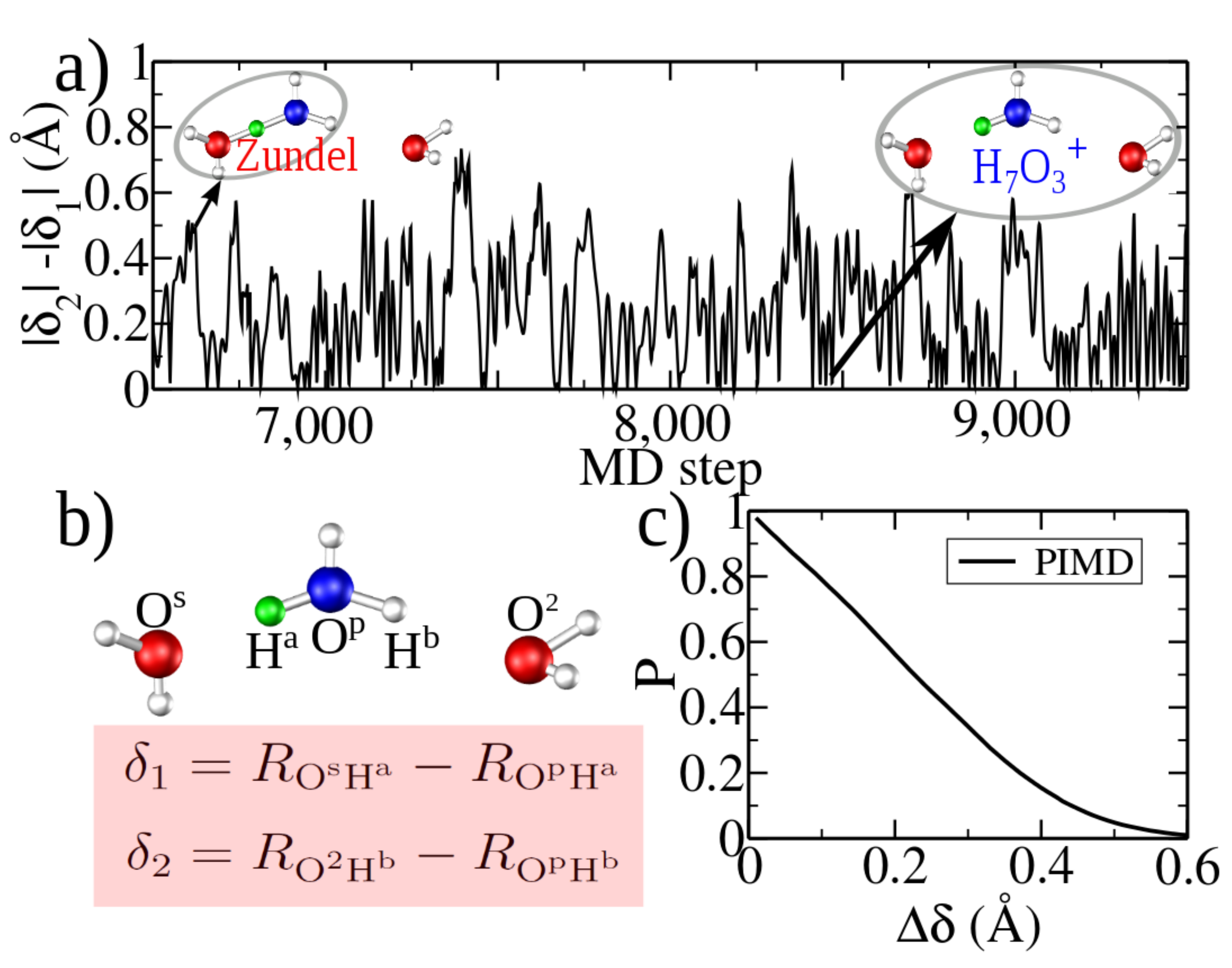}
\caption{\label{figure4}
(a) Evolution of $|\delta_2|-|\delta_1|$, a criterion to distinguish Zundel from the linear H$_7$O$_3^+$
complexes for several thousand time steps in an \textit{ab initio} PIMD simulation of
the hydrated proton in a CNT. Large values of $|\delta_2|-|\delta_1|$ correspond to Zundel-like
species and small values to H$_7$O$_3^+$-like ones. 
(b) Cartoons of an ideal H$_7$O$_3^+$ complex and the definition of $\delta_1$ and $\delta_2$.
To quantify the proportion of Zundel-like
species visited during the simulation, a suitable threshold value of $\Delta \delta$ must
be selected. 
(c) Proportion of Zundel-like species visited during the simulation ($P$) as a function of
this $\Delta \delta$.
We assign all states with $|\delta_2|-|\delta_1|$ larger than $\Delta \delta$ to
Zundel and those with $|\delta_2|-|\delta_1|$ smaller than $\Delta \delta$ to H$_7$O$_3^+$.
The linear dependence indicates that
it is inappropriate to attempt to describe this system
as being comprised mainly of one dominant idealized structure or another. 
}
\end{figure}

For a rigorous quantitative characterization of the interconversion 
between the different hydration states observed,
we use the centroid of the quantum particles and denote
$\delta$ of the most active proton as $\delta_1$.
That of the other O$^{\text{p}}$ hydrogen-bonded proton is 
defined as $\delta_2$ (\ref{figure4}~(b)).
In principle, it is possible to use the difference between $\delta_1$ and $\delta_2$,
\textit{i.e.}, $|\delta_2|-|\delta_1|$ 
(positive by definition) as a means of distinguishing
Zundel from the linear H$_7$O$_3^+$ (which is reported as the transcient intermediate in Ref. \cite{voth})
complexes, since in the Zundel $|\delta_2|-|\delta_1|$ is large whereas
in H$_7$O$_3^+$ $|\delta_2|-|\delta_1|$ is small (\ref{figure4} (a) insets). 
To do this, however, a suitable threshold value ($\Delta\delta$) that distinguishes Zundel from H$_7$O$_3^+$
must be selected, \textit{i.e.}, we need to assign all states with $|\delta_2|-|\delta_1|$ larger 
than $\Delta \delta$ to
Zundel and those with $|\delta_2|-|\delta_1|$ smaller than $\Delta \delta$ to H$_7$O$_3^+$.
Crucially we find here 
that the results obtained for the proportion of Zundel species in the system depend almost
linearly on the value selected for the $\Delta \delta$ threshold. 
This linear dependence is shown in \ref{figure4} (c) and it reveals that it is not useful to attempt to 
describe this system as being comprised
mainly of one dominant idealized structure or another. 
Rather, the fluxional defect
picture for the excess proton as observed in bulk liquid water\cite{marxnat}
should be extended
to this nanoconfined system. 
The evolution of $|\delta_2|-|\delta_1|$ (\ref{figure4} (a)), 
meanwhile, further confirms this conclusion by showing continuous interconversions between states with small and large 
$|\delta_2|-|\delta_1|$.

\section{Discussion}

We now compare our results with other theoretical and experimental studies.
PT both in nanoconfined channels and in bulk liquid water has been 
examined before using various simulation 
methods.\cite{Sagnella, dellago, klein, voth, marxnat, brewer, Koefinger}
However, a detailed analysis
based on a direct comparison of the two classes of system involving fully quantum simulations had been lacking.
From our simulations we find that as a quantum particle the excess proton does
not feel a proton transfer barrier both in bulk water and under confinement.
Rather, the key difference between PT in the two regimes 
lies in solvent reorganization,
where a favorable pre-alignment of water molecules within the tube facilitates facile PT.
States with the quantum excess proton delocalized over several HBs are frequently observed
in both bulk and nanoconfined water, revealing that the fluxional
defect picture of PT in bulk liquid water\cite{marxnat} can be
extended to PT within CNTs.
The delocalized quantum excess proton is consistent with an earlier \textit{ab initio} PIMD 
simulation using a linear water pentamer\cite{klein} confined in a cylindrical external
potential. 
Therefore, we expect that this is a general
feature of 1D PT.

Earlier experimental studies involving water permeability measurements 
have shown a dramatic increase in the diffusion coefficient of H$^+$ compared to other ions when confined within
nanochannels. 
Specifically, it has been shown that the H$^+$ to K$^+$ (Na$^+$) diffusion coefficient ratio
increases from about 9/2 (9/1) in bulk liquid water to 34/2 (34/1) in GA channels.\cite{pomes,roux}
Our understanding that the favorable alignment of water molecules within the nanotube
facilitates facile PT explains in a simple manner this experimental observation.
We note that \textit{ab initio} PIMD does not provide real-time information and 
our discussions are based mainly on statistics and qualitative interpretations of event sequences.
Further investigations within the path-integral scheme, especially those using 
the adiabatic centroid molecular dynamics\cite{cmd1,cmd2,adcmd} and the ring-polymer 
molecular dynamics\cite{rpmd1,rpmd2,rpmd3} methods,
are therefore highly desired for a quantitative description of this issue. 

Finally, the continuous interconversion between ideal hydration models indicates 
that a certain amount of
H$_7$O$_3^+$ states should exist in the nano-confined water chain.
Indeed this appears to be consistent with a recent X-ray study of protons in nanotubular crystals of
H(CHB$_{11}$I$_{11}$)$\cdot$8H$_2$O, 
where similar H$_7$O$_3^+$ structures were reported at low temperature.\cite{stoyanov}
Signatures of the quantum delocalization
in such a 1D system may be detectable with techniques such as electron energy loss
spectroscopy, helium scattering, or neutron scattering. We hope this
work will stimulate such further studies.

\section{Methods}
The MD and PIMD simulations reported here are of the ``Born-Oppenheimer-type'', employing
density-functional theory (DFT).
The VASP\cite{VASP1, VASP2, VASP3, VASP4} code was used, along with our own implementation
of PIMD, which has already been successfully used to study
PT in BaZrO$_3$.\cite{zhang}
For bulk water simulations, 32 water molecules have been employed along with one excess
proton in a 9.87 \AA\text{}$^3$ periodic super-cell.
For PT within the CNT, an excess proton was added
to a one-dimensional periodic hydrogen-bonded water chain.
Six molecules were confined in a 14.8 \AA\text{} long (6,6) CNT with a
radius of 4.1~\AA\text{}. The sensitivity of the results to the
length of the water chain and nanotube were examined with an 8 water molecule chain
in a 19.7~\AA\text{} long nanotube. No significant differences were obtained
compared to the smaller 6 molecule cell.
To account for electron exchange-correlation (xc) interactions,
the Perdew-Burke-Ernzerhof (PBE) xc functional was used.\cite{pbe}
Although PBE does not account for van der Waals (vdW)
dispersion forces, tests with a van der Waals functional,\cite{Dion_2004}
specifically the newly proposed optB88-vdW functional~\cite{Klimes_2010, Klimes_2011}
revealed that the structure and dynamics of the water-filled nanotube is not affected by
the neglect of vdW forces~(Fig.~S1). Similarly the role of exact exchange on the proton
transfer barrier (evaluated with PBE0 and HSE calculations \cite{PBE0, HSE})
was shown to be minimal (Fig.~S1).
Using a time step of 0.5~fs, we report MD (PIMD) results
for 20,000~(15,000) steps, after an equilibration period of 5,000 steps.
32 replicas per nucleus in association with normal-mode coordinates were used in
PIMD to represent the imaginary time path-integral at 300~K,
a temperature controlled by a
Nos\'e-Hoover chain thermostat.~\cite{nose-hoover}
In light of the well-known over-structuring of liquid
water with PBE,\cite{grossman, VandeVondele, Santra}
test calculations were also performed at 330~K
but no significant differences in the PT mechanism compared to the 300 K simulations
were found. Additional computational details and convergence tests are
reported in the supporting materials.

\section{Acknowledgements}
J.C., X.Z.L., Q.F.Z., and E.W. are supported by NSFC. X.Z.L. and A.M. are 
supported by the European Research Council. A.M. is also supported by the 
Royal Society through a Royal Society Wolfson Research Merit Award.
We are grateful for computational resources to the
UK's national high performance
computing service HECToR (for which access was partly obtained via the
UK's Material Chemistry Consortium, EP/F067496).




\footnotesize{
\bibliography{rsc} 
\bibliographystyle{rsc} 
}

\end{document}